FRI-2G.303-1-CCT1-11

# PERFORMANCE EVALUATION OF SUBROUTINES CALL IN PHP [10]

**Yordan Kalmukov, PhD**
Department of Computer Systems and Technologies,
University of Ruse
Tel.: +359 82 888 681
E-mail: jkalmukov@uni-ruse.bg

*Abstract:* One of the most popular and basic principles in programming is the DRY principle (don't repeat yourself). According to it, code duplication should be avoided within a single application. Instead of duplicating it, the code can be exported to/as a subroutine, which can be called as many times as needed and where needed. The same principle is fully adopted and integrated in the object-oriented design as well. It makes the code better structured and more flexible, and significantly facilitates future updates and development of the application. However, there is one problem – cascaded methods calls. Each subroutine call has a price – the code execution time increases and the application performance decreases. The aim of this paper is to conduct a series of experimental analyses to determine how much the performance of a PHP application decreases when the code is exported to subroutines and their subsequent call.

## INTRODUCTION

One of the most popular and basic principles in programming is the DRY principle (don't repeat yourself). According to it, duplication of code in the application should be avoided. Instead of duplicating it, the code can be exported as a subroutine, which can be called as many times as needed and where needed. Behind this principle there is an strong logic - duplication of code creates a lot of problems in maintaining and updating the application, because in case of a possible change, the programmer must necessarily update all copies of the code, otherwise the application will work unreliable.

The same principle is fully adopted and integrated in the object-oriented design as well. According to it, each portion of code that implements a given atomic functionality should be exported into a method or function that can be called as many times as needed and where necessary. This makes the code much better structured and more flexible, and significantly facilitates future updates and development of the application. Because of this, the object-oriented design and the DRY principle, in particular, are de facto global programming standards. However, there is one problem – cascaded method calls. Broader functionality relies on calling methods that implements another smaller functionalities. So, a long chain of methods is formed, in which one method calls a second, the second method a third, and so on. However, each subroutine call has a price – the code execution time increases and the application performance decreases.

The aim of this paper is to conduct a series of experimental analyses to determine how much the performance of a PHP application decreases when the code is exported to subroutines and their subsequent call.

This work is continuation of (Kalmukov, 2025) that aims to determine the best performing loop structure in PHP. Loop structures are also very important in the context of big data analysis and processing.

---

[10] The paper was presented on 24 October 2025 in section "Communication and Computer Technologies" with original title in English: PERFORMANCE EVALUATION OF SUBROUTINES CALL IN PHP

**Related Work**

As known calling a function (or a method) always has some overhead compared to fully inline code, as the interpreter must follow a certain protocol: preserving context, passing arguments, jumping to a new address, returning a value, etc. However that overhead will vary depending on the programming language, its type (compliable or interpretable) and so on. We are interested in determining the function call overhead in PHP particularly, but unfortunately scientific papers related to this issue could not be found on the Internet. However there are PHP enthusiast who performed some experiments and published their results in professional forums like StackOverflow and etc.

Some programmers joke and provide an accurate average percentage - 15.5355% (Johnson, 2013), however their experiments clearly show that the overhead actually varies depending on the operations' complexity.

Many experts state however that rather than thinking which is more efficient, programmers should instead think of which is more maintainable (Planer, 2010).

**Experimental Setup**

The test dataset represents an inverted index, constructed for full text search by using the vector space model (Kalmukov. 2022) within the abstracts of all previous publications of the reviewers of the international conference CompSysTech 2018. The abstracts are extracted from the Semantic Scholar's API (Allen Institute for Artificial Intelligence, 2025).

The structure of the inverted index is as follows:
```
index[<word>]['df'] = df;
index[<word>]['documents'][<docId_1>]= tf_1;
...
index[<word>]['documents'][<docId_m>] = tf_m;
```
where `<word>` is every unique word in the document collection, df is the number of times it appears within the document collection, tf_1 is the number of times it appears in the first document and tf_m is the number of times it appears in the m-th document.

For example:
```
index['rocket']['df'] = 3;
index['rocket']['documents']['doc_11'] = 7;
index['rocket']['documents']['doc_15'] = 2;
index['rocket']['documents']['doc_67'] = 4;
```

The test dataset is a 3-dimensional associative array where the first dimension index is a string index, which in this case is every unique word in the document collection. The collection contains **4573 abstracts** and **21624 semantically significant unique words**. So, the test array has 21624 elements, which in turn are other bi-dimensional arrays of multiple elements.

To *evaluate how much the performance is affected* in both cases – built-in/inline code vs. subroutine/function call – operations of different complexity should be implemented and executed. Having the context (full text document search) in mind, probably the most reasonable operations to implement are *calculating term weights* based on two widely utilized term-weighting models (Kalmukov, 2020) – the basic lnc.ltc model (1) and the Robertson BM 25 model (2).

Basic tf-idf model:

$$w_{i,j} = \left(1 + \log(tf_{i,j})\right) * \log\left(\frac{d}{df_i}\right) \qquad (1)$$

where:
$w_{i,j}$ – the weight of the $i$-th term in the $j$-th document.
$tf_{i,j}$ – the number of times the term $t_i$ appears in the $j$-th document.
$d$ – the number of documents in the entire document collection.
$df_i$ – number of documents that contain $t_i$.

$tf_{i,j}$ and $df_i$ are taken directly from the index (input dataset).

Robertson's BM 25 model:

$$w_{i,j} = \frac{(k_1 + 1)(tf_{i,j})}{k_1\left(1 - s + s\frac{dl(d_j)}{avdl}\right) + tf_{i,j}} \log(1 + \frac{d}{df_i}) \qquad (2)$$

where:
$k_1$ – a constant set to 1.2 by default. It could be tuned, depending on the data being processed.
$b$ – slope parameter. $\boldsymbol{b \in [0, 1]}$. By default b = 0.2.
$dl(d_j)$ – length of the current, $j$-th, document.
$avdl$ – average document length within the entire document collection.

Since the two models implement different number of operations, they have different time complexity. Therefore, they are ideal for testing if the function call overhead is constant for all cases or it depends on the number and complexity of the operations included in the code/subroutine.

Experiments are performed on a desktop computer having an AMD Ryzen 3 1300X (4 core, no multithreading) CPU, 16 GB of RAM DDR 4 single channel, Windows 10 22H2, PHP v.7.3.9 running as a module of Apache/2.4.41 HTTP server.

**Results and Discussions**

Experiment 1: Calculating term weights of all 21 624 unique words by using the basic tf-idf model (formula 1)

Inline code:
```
$w_ij = (1+log($tf)) * log($docCount/$df+1);
```

Function call:
```
$w_ij = calculateDocumentWeight($tf, $df, $docCount);
function calculateDocumentWeight($tf, $df, $d) {
     return (1+log($tf)) * log($d/$df+1);
}
```

All experiments are repeated 3 times in order to reduce the influence of the co-called random/uncontrollable factors such as the system's current load from other processes. Then the execution time is averaged and the average value is taken into account in the subsequent analyses.

Table 1. Execution time for calculating term weights by using basic tf-idf model (1): inline code vs. code exported and called from a function.

|  | *Execution time in microseconds* | | | | |
|---|---|---|---|---|---|
|  | Repetition 1 | Repetition 2 | Repetition 3 | Average |  |
| **Basic tf-idf** | | | | | |
| Inline code | 26.782 | 29.032 | 27.5731 | **27.7957** | |
| Function call | 41.4691 | 42.3369 | 43.0271 | **42.2777** | **52%** overhead |

It can be reasonably assumed that the overhead will not always be fixed, but depends on how many and how complex operations are performed within the function / the inline code. The assumption is tested in Experiment 2.

Experiment 2: Calculating term weights of all 21 624 unique words by using the Robertson's BM 25 model (formula 2)

Inline code:
```
$k1 = 1.2;
$s = 0.2;
$mul = (($k1+1)*$tf) / ($k1*(1-$s + $s*
($documentInfo[$docId]['meaningfullWordsCount']/$averageDocLen))+$tf);
$idf = log(1 + ($docCount/$df));
$w_ij_BM25 = $mul*$idf;
```

Function call:
```
$w_ij_BM25 = calculateDocumentWeight_BM25($tf, $df, $docCount,
$documentInfo[$docId]['meaningfullWordsCount'], $averageDocLen);

function calculateDocumentWeight_BM25($tf, $df, $d, $dl, $avdl) {
    $k1 = 1.2;
    $s = 0.2;
    $mul = (($k1+1)*$tf) / ($k1*(1-$s + $s*($dl/$avdl)) + $tf);
    $idf = log(1 + ($d/$df));
    return $mul*$idf;
}
```

Table 2. Execution time for calculating term weights by using Robertson's BM 25 model (2): inline code vs. code exported and called from a function.

|  | *Execution time in microseconds* | | | | |
|---|---|---|---|---|---|
|  | Repetition 1 | Repetition 2 | Repetition 3 | Average |  |
| **BM 25** | | | | | |
| Inline code | 45.5258 | 46.7951 | 44.3270 | **45.5493** | |
| Function call | 65.3789 | 66.2060 | 64.2560 | **65.2803** | **43%** overhead |

As expected, with increasing the number and the complexity of the operations within the code, the overhead percentage decreases. The reason is that the cost of calling a subroutine seems to be fixed, so increasing the number of operations within the code (and/or their complexity) lowers the relative percentage of that cost in respect to the time needed for execution of the "useful payload" (the useful part of the code that implements some functionality).

Although this observation is relies is backed by strong logic, it could be further verified by increasing the number of operations even more (experiment 3).

Experiment 3: Calculating term weights of all 21 624 unique words by using the modified Robertson's BM 25 model.

The model is modified by multiplying both terms (tf and idf) with a constant (for example 100), then dividing each multiplication by the same constant. This will not change the calculated term weights, but adds more operations.

Inline code:
```
...
$w_ij_BM25 = (($mul*100)/100) * (($idf*100)/100);
```

Function call:
```
...
return (($mul*100)/100) * (($idf*100)/100);
```

Table 3. Execution time for calculating term weights by using modified Robertson's BM 25 model: inline code vs. code exported and called from a function.

|  | *Execution time in microseconds* | | | | |
|---|---|---|---|---|---|
|  | Repetition 1 | Repetition 2 | Repetition 3 | Average |  |
| **Modified BM 25** |  |  |  |  |  |
| Inline code | 59.2041 | 56.5951 | 56.8681 | **57.5558** |  |
| Function call | 72.8710 | 75.3500 | 76.3669 | **74.8626** | **30%** overhead |

Indeed, adding more operations decreased the relative function call overhead.

Another interesting experiment would be to check how exactly the overall execution time depends on the size of the input dataset. The dependency should be linear, but could not be stated for sure unless it is experimentally proven.

Experiment 4: Dependency of the execution time on the size of the input dataset

For the purpose of this experiment the inverted index is replicated many times (1, 5, 10, 15 and 20 times), then all replicas are united in a single array (with duplicated elements obviously). Since term weights are not cached anywhere, the element duplication should not influence the linearity of the execution time.

Table 4. Execution time for calculating term weights by using the basic tf-idf model (1): inline code vs. code exported and called from a function. Determining the dependency of the execution time on the size of the input dataset.

|  |  | *Execution time in microseconds* |  |  |  |  |
|---|---|---|---|---|---|---|
|  |  | Repetition 1 | Repetition 2 | Repetition 3 | Average | Overhead |
| Original index, 21 624 words |  |  |  |  |  |  |
|  | Inline code | 47.384 | 48.7361 | 44.96 | 47.0267 |  |
|  | Function call | 69.9211 | 72.5639 | 66.709 | 69.7313 | 48.28 % |
| 5 copies of the index, 108 120 words |  |  |  |  |  |  |
|  | Inline code | 256.763 | 242.264 | 236.9559 | 245.3276 |  |
|  | Function call | 362.6449 | 369.0159 | 342.0999 | 357.9202 | 45.89 % |
| 10 copies of the index, 216 240 words |  |  |  |  |  |  |
|  | Inline code | 461.5929 | 462.5301 | 464.2661 | 462.7964 |  |
|  | Function call | 682.039 | 668.529 | 673.7299 | 674.7660 | 45.80 % |
| 15 copies of the index, 324 360 words |  |  |  |  |  |  |
|  | Inline code | 652.5021 | 676.9481 | 672.0302 | 667.1601 |  |
|  | Function call | 953.3739 | 981.6539 | 973.3469 | 969.4582 | 45.31 % |
| 20 copies of the index, 432 480 words |  |  |  |  |  |  |
|  | Inline code | 899.018 | 915.719 | 886.435 | 900.3907 |  |
|  | Function call | 1305.4609 | 1302.7701 | 1320.523 | 1309.5847 | 45.45 % |

The tabular data shows linear dependency of execution time on the size of input dataset. This could be visually confirmed on figure 1.

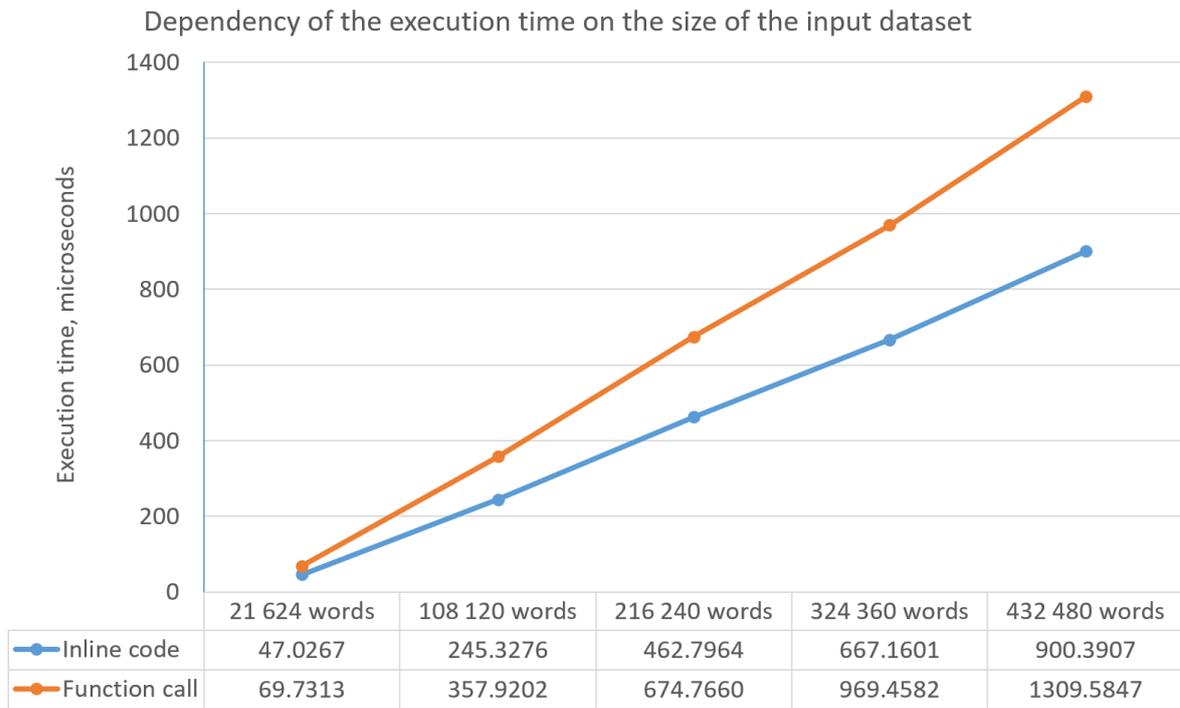

Fig. 1. Dependency of the execution time on the size of input data.

## CONCLUSION

After performing dozens of experiments, it could be concluded that:

1. Exporting code to subroutines and their subsequent call significantly affects the execution time of the application, especially when it comes to exporting a small number of elementary operations.

2. In the case of a lightweight web application that does not perform heavy processing of a large amount of data, flexibility and good structure of the application are more important than performance, i.e. the DRY principle and object-oriented design should be followed.

3. In web applications that process a large amount of data, heavy operations could be moved to subroutines, but exporting a small number of elementary operations is not justified, since the time to call the subroutine may even exceed the time to run the useful code.

## ACKNOWLEDGEMENTS

This paper is supported by project 2025-FEEA-02 "Application of AI in various professional fields: opportunities, challenges and ethical issues", funded by the Research Fund of the "Angel Kanchev" University of Ruse, Bulgaria.